# Molecular Dynamics Simulations of Membrane Selectivity of Star Peptides Across Different Bacterial and Mammalian Bilipids

Amal Jayawardena[1], Andrew Hung[2], Greg Qiao[3], Neil O'Brien-Simpson[4], Elnaz Hajizadeh[1*]

[1] Soft Matter Informatics Research Group, Department of Mechanical Engineering, Faculty of Engineering and Information Technology, University of Melbourne, Parkville, VIC, 3010, Australia

[2] School of Science, STEM College, RMIT University, VIC, 3001, Australia

[3] Department of Chemical Engineering, Faculty of Engineering and Information Technology, University of Melbourne, Parkville, VIC, 3010, Australia

[4] ACTV Research Group, Division of Basic and Clinical Oral Sciences, The Melbourne Dental School, Royal Dental Hospital, The University of Melbourne, Melbourne, Victoria 3010, Australia

*Corresponding Author: ellie.hajizadeh@unimelb.edu.au

## Abstract

Star-shaped peptides a.k.a. structurally nanoengineered antimicrobial peptide polymers (SNAPPs) are emerging as promising selective agents against bacterial membranes. In this study, we used all atom molecular dynamics simulation techniques to investigate the interaction of a promising cationic SNAPP architecture (Alt-SNAPP with 8 arms made of alternating lysine and valine residues) with modelled Gram-negative, Gram-positive, mammalian, and red blood cell membranes. Alt-SNAPP exhibited rapid and stable binding to bacterial membranes, driven by electrostatic interactions with anionic lipids such as phosphatidylglycerol (PG) and cardiolipin (CL), and supported by membrane fluidity. In contrast, mammalian and red blood cell membranes, enriched in zwitterionic lipids and cholesterol, resisted peptide association entirely. Analyses of center of mass distance, partial density, hydrogen bonding, and interaction energy confirmed that SNAPP remains fully excluded from host like membranes while forming stable, multivalent interactions with bacterial bilayers. These findings provide mechanistic insight into SNAPP's membrane selectivity and offer a molecular framework for designing next generation antimicrobial polymers with minimal off target toxicity.

## 1. Introduction

The growing crisis of antimicrobial resistance (AMR) continues to pose a major challenge to modern medicine, with traditional antibiotics becoming increasingly ineffective against a broad range of pathogenic bacteria[1-6]. This urgency has accelerated the search for alternative antimicrobial strategies, particularly those that can bypass common resistance mechanisms. Among these, antimicrobial peptides (AMPs) have gained widespread interest due to their ability to physically disrupt bacterial membranes rather than relying on intracellular targets, thereby reducing the potential for resistance development[7, 8].

Structurally nanoengineered antimicrobial peptide polymers (SNAPPs) represent a next-generation class of synthetic antimicrobials that build upon the membrane targeting mechanism of classical AMPs[9-13]. These star shaped polymers are designed to engage in multivalent interactions with bacterial membranes, amplifying membrane disruption through a combination of electrostatic attraction and hydrophobic anchoring[14]. This unique structural arrangement not only enhances bactericidal efficacy but also improves stability and resistance to enzymatic degradation.

A crucial requirement for therapeutic application of SNAPPs and similar membrane-active peptides is achieving selective targeting of bacterial membranes without adversely affecting mammalian cells. Bacterial membranes typically contain a high proportion of negatively charged phospholipids such as phosphatidylglycerol (PG) and cardiolipin, whereas mammalian membranes are rich in zwitterionic phosphatidylcholine (PC) and often contain cholesterol[15-19]. These compositional differences form the molecular basis for selective interaction[17, 20-22]. However, the physical similarity between the lipid bilayers of bacterial and mammalian cells makes complete discrimination challenging. Membrane selectivity is therefore a key design criterion in AMP development to avoid cytotoxicity and hemolysis[20].

Experimental work has demonstrated that SNAPPs exhibit minimal toxicity toward mammalian cells. Studies involving hemolysis assays and cell viability measurements have shown that SNAPPs remain non-disruptive to mammalian membranes at antimicrobial concentrations, suggesting intrinsic selectivity toward bacterial bilayers[9-11]. However, the molecular determinants of this selectivity, and the extent to which chemical modifications can enhance or compromise it, remain poorly understood.

To investigate the molecular basis of this observed selectivity, we employed all-atom molecular dynamics simulations to study the interactions of a model SNAPP featuring a star-shaped architecture. As shown in Figure 1e, this polypeptide consists of eight arms extending from a central hydrophobic core. Each arm follows a repeating lysine and valine sequence (KKVKKVKKVKKV). Due to this alternating residue pattern, we refer to the construct as Alt-SNAPP. SNAPPs can adopt a range of configurations by varying the number of arms, the amino acid composition within each arm, and the arm length. Experimental studies[9-11] have shown that the eight-arm variant exhibits good performance. Consistent with these findings, our previous computational studies have also employed the eight-arm SNAPP model[13, 14]. This model captures the key features of experimentally synthesized SNAPPs[9] and enables a systematic evaluation of peptide membrane interactions at atomic resolution.

The Alt-SNAPP model was placed above a series of biologically relevant lipid bilayers representing Gram-negative, Gram-positive, mammalian and red blood cell membranes as shown in Figures 1a, 1b, 1c, and 1d. These membrane systems were selected to reflect the compositional diversity encountered in bacterial versus mammalian cells and to allow direct comparison of interaction behaviors[15-19].

This manuscript is organized as follows. Section 2 describes the methodology used to construct the Alt-SNAPP and membrane systems and outlines the simulation parameters and analysis techniques employed. These include all-atom molecular dynamics simulations, center-of-mass (COM) distance calculations of the center-of-mass (COM) distance of Alt-SNAPP from the upper leaflet of the bilayer along the membrane normal (z-axis), partial density profiling, hydrogen bond analysis, and decomposition of interaction energies. Section 3 presents the results and discussion and is structured into four main parts. The Subsection 3.1 examines the time taken by Alt-SNAPP to approach the membrane surface by tracking the center-of-mass (COM) distance relative to the upper leaflet of the bilayer. The Subsection 3.2 presents partial density profiles to assess the spatial distribution of Alt-SNAPP relative to the membrane. In the Subsection 3.3, the number of hydrogen bonds formed between the peptide and lipid headgroups are quantified to evaluate specific polar interactions. The Subsection 3.4 analyzes the nonbonded interaction energies between Alt-SNAPP and the membrane, further resolved by amino acid residues and lipid types, to identify the driving forces of membrane association. Finally, Section 4 provides the conclusion, summarizing the key findings and highlighting their relevance to the design of selective, membrane-active antimicrobial polymers.

## 2. Methodology

All atom molecular dynamics simulations were performed using the GROMACS 2023.5 software package[23]. The CHARMM36 force field was employed based on its demonstrated accuracy in reproducing lipid bilayer properties and peptide–membrane interactions[24]. The Alt-SNAPP structure consisted of eight arms, each featuring an alternating sequence of lysine and valine residues in a 2:1 ratio[14], as illustrated in Figure 1e.

The Alt-SNAPP structure was built using a modular approach. Initially, individual arms were modelled using Avogadro molecular modelling software[25], which allowed for precise sequence control and the specification of initial geometries. Each arm was constructed in an α-helical conformation to reflect their experimentally observed behaviour in aqueous or membrane environments[9-11]. A central hydrophobic core was then generated using Packmol[26], approximated as a spherical cluster of 400 carbon atoms to mimic the steric bulk and hydrophobic nature of the polymer core. The eight arms were covalently bonded to core atoms via their C-terminal carbonyl groups, with fixed bond lengths of 2.5 Å, yielding a symmetrical, star-shaped SNAPP.

Experimentally, SNAPPs are synthesized through ring-opening polymerization of lysine and valine N-carboxyanhydrides (NCAs), initiated by PAMAM dendrimers bearing terminal amine groups. This synthetic method ensures covalent linkage of the C-terminus of lysine residues to the core, creating the multivalent, star-like architecture[9].

Table 1 - Lipid composition of the bacterial and mammalian membrane models used in this study. Abbreviations: POPE, 1-palmitoyl-2-oleoyl-sn-glycero-3-phosphoethanolamine; POPG, 1-palmitoyl-2-oleoyl-sn-glycero-3-phosphoglycerol; PVCL2, 1,1-palmitoyl-2,2-vacenoyl cardiolipin; POPC, 1-palmitoyl-2-oleoyl-phosphatidylcholine; PSM, palmitoyl-sphingomyelin; Chol, cholesterol; POPS, 1-palmitoyl-2-oleoyl-phosphatidylserine. The visual representations of the bilayer systems, along with their corresponding lipid colour codes, are shown in Figures 1a–1d.

| Type of Membrane | Lipid composition | |
|---|---|---|
| **Modelled Gram-negative** | POPE (Blue) – 70% <br> POPG (Pink) – 30% | |
| **Modelled Gram-positive** | POPG (Pink) - 60% <br> PVCL2 (Grey) – 40% | |
| **Modelled Mammalian** | POPC (Orange) - 100% | |
| **Red Blood Cells (RBC)** | Exo <br><br> PSM (Green) – 43% <br> POPC (Orange) - 40% <br> CHOL (Black) – 11% <br> POPE (Blue) – 3% <br> POPS (Purple) – 3% | Endo <br><br> PSM (Green) – 4% <br> POPC (Orange) - 22% <br> CHOL (Black) – 11% <br> POPE (Blue) – 38% <br> POPS (Purple) – 25% |

Atomistic bilayer membranes were constructed using the CHARMM-GUI Membrane Builder[27, 28]. As detailed in Table 1, each bilayer was constructed using lipid compositions representative of Gram-negative, Gram-positive, mammalian, and red blood cell membranes[15-19]. These compositions were applied symmetrically across both leaflets, except in the case of red blood cells, where distinct lipid compositions were used for the inner and outer leaflets to reflect physiological asymmetry (Table 1). The SNAPP molecules were manually positioned using VMD[29] such that the center of mass of Alt-SNAPP was placed 8 nm above the upper leaflet of the bilayer, minimizing orientation bias at the start of the simulation in accordance with previously established protocols as shown in Figures 1a, 1b, 1c and 1d[30-32]. Potassium and Chloride ions were added to neutralize the system and approximate a physiological salt concentration of 150 mM. The full details of the simulation box dimension, lipid composition, water content, ion counts, and total simulation times for all systems are summarized in Table 2.

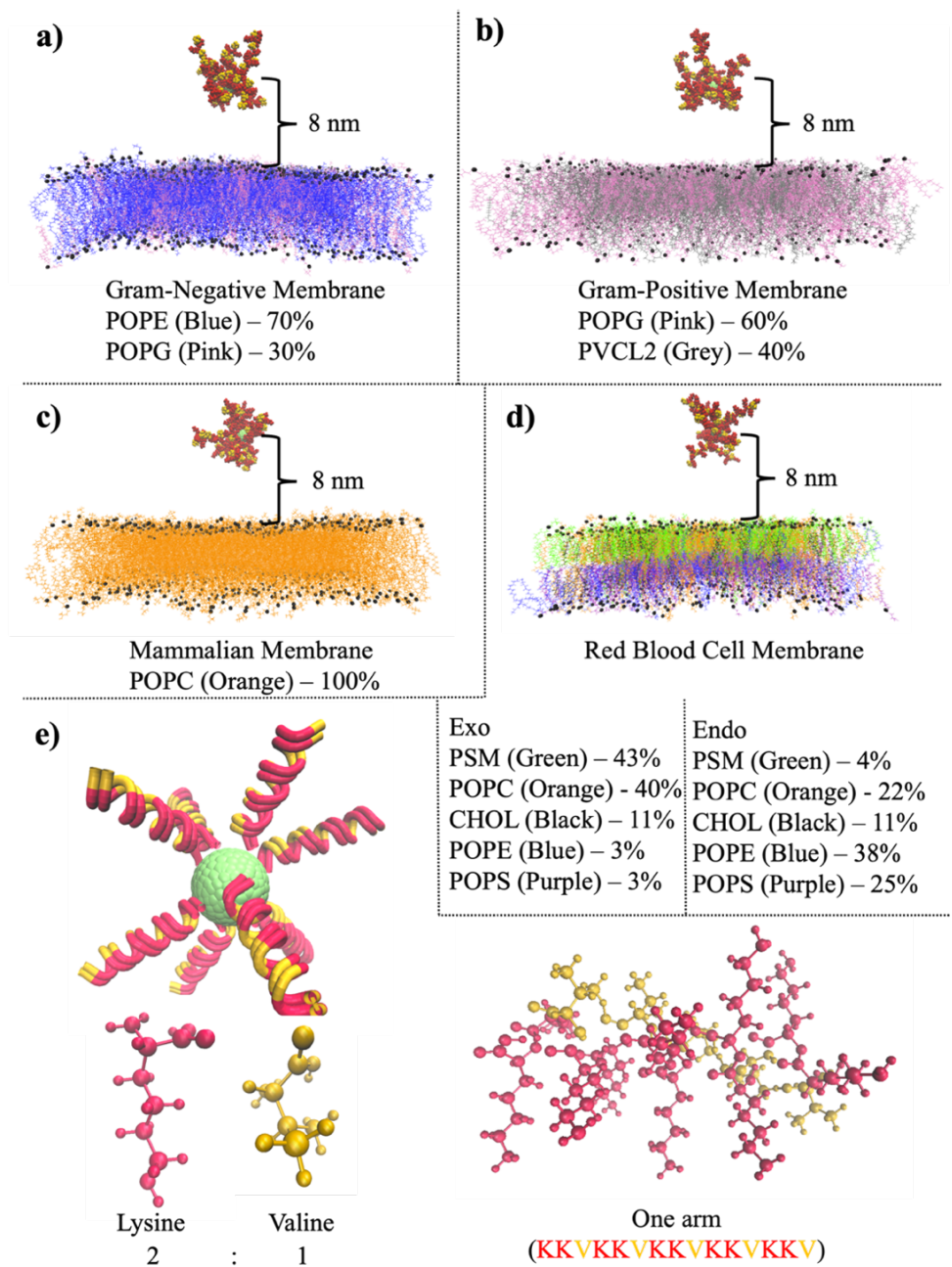

*Figure 1 - Initial configuration of Alt-SNAPP positioned approximately 8 nm above the surface of each modelled membrane system: a) modelled Gram-negative membrane composed of POPE (blue) and POPG (pink), b) modelled Gram-positive membrane composed of POPG (pink) and PVCL2 (grey), c) modelled mammalian membrane composed entirely of POPC (orange), and d) modelled red blood cell (RBC) membrane with an asymmetric distribution of POPC (orange), PSM (green), CHOL (black), POPE (blue), and POPS (purple); lipid types are coloured according to their identity, and phosphate headgroups are shown in black in all membrane system. e) Structural representation of the star shaped Alt-SNAPP molecule, consisting of eight arms with a 2:1 repeating lysine (K) and valine (V) sequence in alternating order.*

Energy minimization was carried out using the steepest descent algorithm until the maximum force in the system was less than 1000 kJ·mol$^{-1}$·nm$^{-1}$. The systems were then equilibrated for 50 ns under constant pressure and temperature (NPT ensemble). The LINCS algorithm was used to constrain all bonds involving hydrogen atoms[33], and electrostatics were treated using the Particle Mesh Ewald (PME) method[34]. During equilibration, the V-rescale thermostat[35] and C-rescale barostat[36] were used to control temperature and pressure, respectively. The temperature was maintained at 303.15 K with a coupling constant of 1 ps, while pressure was controlled at 1 bar using semi-isotropic pressure coupling with a time constant of 5 ps and compressibility 4.5 × 10$^{-5}$ kJ$^{-1}$·mol·nm³. Production simulations were also conducted using the V-rescale thermostat and the C-rescale barostat to maintain the same conditions.

*Table 2 - System composition for each SNAPP membrane system studied. The table summarizes the box dimensions, number of lipid molecules (by type), total water molecules (TIP3), ion counts (K$^+$ as POT and Cl$^-$ as CLA), and total simulation time for each membrane system interacting with the Alt-SNAPP.*

| Simulation | Box size (nm) | Number of lipid molecules | Number of water molecules | Number of ions | Simulation time (ns) |
|---|---|---|---|---|---|
| Gram-negative membrane vs Alt-SNAPP | 19.78199 19.78199 39.72085 | POPE - 938 POPG - 402 | TIP3 - 466072 | POT - 1769 CLA - 1447 | 500 |
| Gram-Positive membrane vs Alt-SNAPP | 20.07707 20.07707 38.44720 | POPG - 540 PVCL2 - 360 | TIP3 - 467460 | POT – 2630 CLA - 1450 | 500 |
| Mammalian membrane vs Alt-SNAPP | 19.33693 19.33693 41.18354 | POPC - 1172 | TIP3 - 465651 | POT - 1446 CLA - 1526 | 500 |
| Red blood cell membrane vs Alt-SNAPP | 18.85945 18.85945 40.35068 | POPC – 434 PSM – 329 POPS – 196 CHL – 154 POPE - 287 | TIP3 - 478965 | POT – 1611 CLA - 1495 | 500 |

**Time to Contact Analysis via Center of Mass Distance**

The time required for Alt-SNAPP to reach the membrane surface was assessed by calculating the center of mass (COM) distance of the Alt-SNAPP from the top leaflet of the bilayer along the membrane normal (z-axis) as a function of time. The top leaflet was defined based on the phosphate atoms of the upper bilayer, which served as a structural reference for the membrane surface. This analysis was conducted using the gmx distance module in GROMACS, with custom index groups used to define Alt-SNAPP and membrane components.

**Partial Density Profiles Analysis**

The spatial distribution of Alt-SNAPP along the membrane normal was analysed by computing partial mass density profiles using the gmx density tool in GROMACS. The simulation box was divided into narrow slabs along the z-axis, and the average mass density was calculated for Alt-SNAPP. This method enabled visualization of the relative position and localization of Alt-SNAPP across different membrane environments, providing insight into membrane association and selectivity.

**Hydrogen Bonding Analysis**

To investigate residue-specific and lipid-specific polar interactions, hydrogen bonds formed between Alt-SNAPP and membrane lipids were analysed using the hydrogen bond analysis plugin in VMD[29]. The analysis was performed separately for each lipid species, allowing quantification of hydrogen bonds formed by lysine and valine residues of Alt-SNAPP with individual lipid types. Hydrogen bonds were resolved by both lipid identity and residue type to assess how different components contributed to the membrane association. This approach provided a detailed understanding of the specific roles of lysine, valine, and distinct lipid headgroups in mediating polar contacts at the membrane interface.

**Interaction Potential Energy Analysis**

Interaction potential energies between the Alt-SNAPP and membrane components were calculated using the gmx energy module in GROMACS. These short-range nonbonded interactions were computed within the defined cutoff distance and separated into Coulombic (electrostatic) and van der Waals (Lennard Jones) components, as defined by the force field. To better understand the molecular forces that drive Alt-SNAPP interactions with membranes, the analysis was extended beyond the total interaction energy.

Energies were further resolved by amino acid type, specifically lysine and valine, and by individual lipid species within each membrane system. This residue-level and lipid-level resolution allowed us to quantify the contributions of positively charged and hydrophobic residues to the interaction energy with each lipid type. For each interaction pair, Coulombic, van der Waals, and total interaction energies were calculated and plotted separately. This approach enabled a detailed comparison of the molecular interactions that influence Alt-SNAPP's selectivity toward bacterial and mammalian membranes.

## 3. Results and Discussion

### 3.1. Time to Membrane Contact for Alt-SNAPP with Gram-negative, Gram-positive, Mammalian, and Red Blood Cell Membranes

The molecular dynamics simulations revealed distinct patterns of Alt-SNAPP interaction across the different membrane systems, primarily governed by lipid composition and bilayer properties. These trends are consistent with experimental observations, indicating that SNAPPs preferentially target bacterial membranes, while sparing mammalian and erythrocyte membranes, a feature critical for their therapeutic application[9-11].

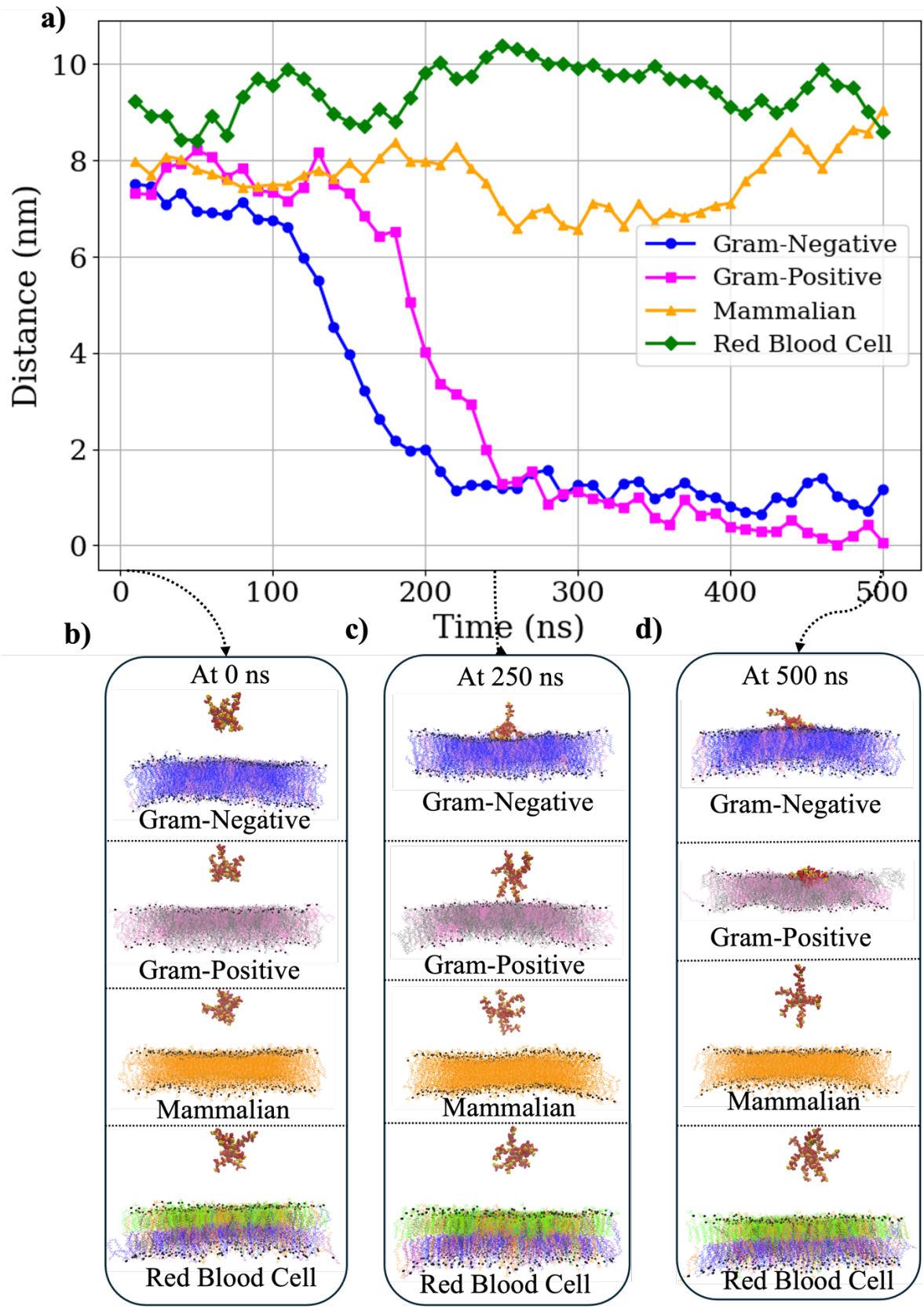

Figure 2 - Time evolution of Alt-SNAPP approach toward the membrane surface. a) Center of mass (COM) distance of the Alt-SNAPP from the bilayer surface over time, starting from an initial separation of approximately 8 nm; the COM distance reflects the vertical displacement between the peptide and the phosphate plane of the upper leaflet. b) Representative snapshots at 0 ns showing the initial placement of Alt-SNAPP above each modelled membrane system. c) Snapshots at 250 ns illustrating the intermediate approach toward the bilayer surface.

*d) Snapshots at 500 ns showing the final position of Alt-SNAPP relative to each membrane. All structures were visualized using VMD.*

In the modelled Gram-negative membrane composed of 70 percent phosphatidylethanolamine (POPE) and 30 percent phosphatidylglycerol (POPG), Alt-SNAPP rapidly approached the bilayer, with the center of mass (COM) distance decreasing from 8 nm to 2 nm within the first 200 ns, as shown by the blue curve in Figure 2a. From 200 ns to 500 ns, the peptide continued to move toward the membrane surface, eventually reaching the upper leaflet, as illustrated in Figures 2a through 2d. This behaviour is likely driven by strong electrostatic attraction between the cationic branches of SNAPP and the anionic surface provided by POPG, further facilitated by the loosely packed, fluid structure of POPE as shown in Figure 3[37]. The observed rapid binding aligns with experimental data showing fast membrane disruption in Gram-negative bacteria upon SNAPP exposure[9], and with findings that polycationic antimicrobial agents favour interaction with disordered, negatively charged bilayers[38].

In contrast, the Gram-positive membrane model, composed of 60 percent phosphatidylglycerol (POPG) and 40 percent PVCL2, a doubly anionic cardiolipin species, exhibited a slower initial interaction with Alt-SNAPP as shown in Figures 2a magenta line. The center of mass (COM) distance of the Alt-SNAPP from the Gram-positive membrane remained above 7 nm for the first 150 ns, followed by a gradual approach that reached approximately 4 nm by 200 ns. In comparison, Alt-SNAPP reached the 2 nm mark by 200 ns in the Gram-negative membrane. As shown quantitatively in Figure 2a and qualitatively in Figure 2c, Alt-SNAPP positioned itself farther from the bilayer headgroups in the Gram-positive system during the early phase of the simulation. Although the initial association was delayed relative to the Gram-negative membrane, the final insertion depth was slightly greater in the Gram-positive system, suggesting a more stable and deeper engagement once binding occurred, as illustrated in Figures 2a and 2d. This observation suggests that the timing and depth of interaction are influenced by different membrane molecular features.

While both membranes share a high surface charge due to the presence of anionic lipids, the delayed interaction in the Gram-positive system may result from the structural effects of cardiolipin[39]. As a tetra acyl phospholipid with a conical geometry and a net charge of -2, cardiolipin is known to promote tighter lipid packing, increase membrane order, and reduce lateral lipid mobility. These characteristics create a more rigid and organized environment that may initially resist peptide penetration[40-42]. However, once Alt-SNAPP engages with the bilayer, its interaction appears more persistent and penetrative than in the Gram-negative system.

Although the peptidoglycan matrix of Gram-positive bacteria was not included in the model, these results show that lipid composition alone, particularly the presence of cardiolipin, can significantly affect both the rate and extent of peptide membrane association. This behaviour aligns with experimental findings that cardiolipin enriched membranes are less susceptible to immediate disruption but can be targeted over longer timescales by cationic antimicrobial agents [42, 43].

In stark contrast, the modelled mammalian membrane, made entirely of phosphatidylcholine (POPC), demonstrated minimal SNAPP interaction as shown in Figure 2a orange line. POPC is a zwitterionic lipid with no net charge, and the membrane lacks anionic components or structural defects that would enable electrostatic attraction or insertion. The COM distance remained stable throughout the 500 ns, indicating that SNAPP failed to associate with the bilayer. This result is consistent with biophysical studies showing that POPC only membranes resist peptide insertion due to their high lipid order and neutral headgroup chemistry[44]. Importantly, this aligns with *in vitro* data showing that SNAPPs exhibit minimal cytotoxicity to mammalian cells[9-11], a feature directly supported by the lack of observed membrane interaction in our simulations.

The red blood cell (RBC) membrane model was composed of two asymmetric leaflets: the exoplasmic leaflet contained 43 percent PSM (palmitoyl sphingomyelin), 40 percent POPC, 11 percent cholesterol (CHOL), 3 percent POPE, and 3 percent phosphatidylserine (POPS), while the endoplasmic leaflet consisted of 4 percent PSM, 22 percent POPC, 11 percent CHOL, 38 percent POPE, and 25 percent POPS. This composition reflects the well-documented asymmetry and cholesterol enrichment of erythrocyte membranes[45]. Throughout the simulation, SNAPP showed no attempt to approach, with the COM distance remaining above 9 nm as shown in Figure 2a green line. High amount of sphingomyelin and cholesterol in the outer leaflet result in a tightly packed, ordered domains that strongly resist peptide adsorption and insertion. To quantitatively assess this effect, Figure 3 presents the evolution of the area per lipid over a 500 ns simulation period. The results reveal a progressive reduction in the headgroup area within the red blood cell membrane compared with the Gram-negative membrane, highlighting the enhanced lipid packing density in the former. It is also important to note that although periodic boundary conditions theoretically allow the peptide to drift toward the lower (intracellular) leaflet, this did not occur, as SNAPP consistently remained closer to the upper leaflet throughout the trajectory as shown in Figure 2. These results mirror experimental findings that SNAPPs exhibit minimal haemolytic activity, even at bactericidal concentrations[9] and emphasize the biophysical barrier that cholesterol and sphingomyelin rich membranes pose to antimicrobial polymers[46, 47].

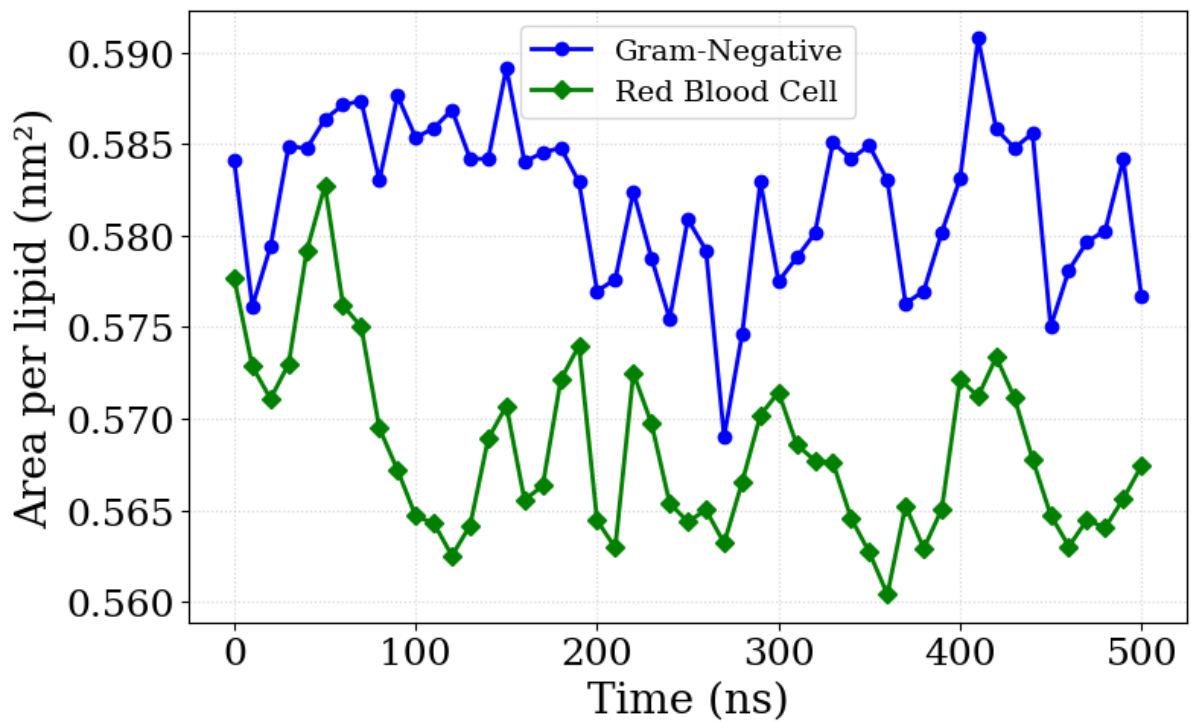

*Figure 3 - Comparison of area per lipid between red blood cell and Gram-negative membranes over 500 ns of simulation.*

Collectively, the molecular dynamics simulation results establish a clear mechanistic explanation for the membrane selectivity exhibited by Alt-SNAPP. The degree of SNAPP interaction is strongly influenced by the composition of the lipid bilayer, particularly the presence of anionic lipids such as phosphatidylglycerol and phosphatidylethanolamine, and the absence of lipid species that promote ordered membrane domains, including sphingomyelin and cholesterol. In bacterial membranes enriched with negatively charged and fluid lipid components, SNAPP readily associates and remains stably embedded near the membrane interface. In contrast, mammalian and red blood cell membranes, which are composed primarily of zwitterionic lipids such as phosphatidylcholine and sphingomyelin and contain high levels of cholesterol, exhibit minimal to no interaction with SNAPP.

To further characterize the structural basis of the observed selectivity, we conducted partial density analyses of Alt-SNAPP and the membrane along the bilayer normal. This analysis allowed us to visualize the spatial distribution of the peptide relative to the membrane and to quantify the degree of insertion and overlap in each system.

### 3.2. Spatial Distribution and Depth of SNAPP Association Across Membrane Types

To quantitatively assess the spatial arrangement of SNAPP in each membrane environment, we computed normalized partial density profiles along the bilayer normal for both SNAPP and the surrounding membrane lipids. These profiles provide insight into the extent of vertical penetration and alignment relative to the bilayer interface and complement the center-of-mass (COM) distance analysis presented earlier.

In the Gram-negative membrane system, the SNAPP density peak partially overlaps with the headgroup region of the lipid bilayer as shown in Figure 4a. This spatial coincidence indicates that SNAPP is positioned at or just above the phosphate-rich zone, consistent with surface binding and shallow insertion. The interaction is likely driven by electrostatic attraction between the multivalent cationic SNAPP and the anionic POPG, while POPE contributes to bilayer fluidity and negative intrinsic curvature that may further promote binding[37, 48-50].

In the case of Gram-positive membrane, Alt-SNAPP exhibits a similar overall profile, with its peak marginally more inserted than in the Gram-negative system as shown in Figure 4b. While the difference is modest, it reflects a slightly deeper engagement with the bilayer, despite the presence of more ordered lipid domains imposed by PVCL2. This finding suggests that higher anionic lipid content may enhance electrostatic anchoring, enabling Alt-SNAPP to access regions closer to the upper acyl chain zone of the bilayer. Previous studies[51, 52] have reported similar trends, demonstrating that deeper peptide insertion can occur in membranes with increased negative surface charge density, even in the presence of elevated bilayer stiffness. This highlights the dominant role of electrostatics in guiding the initial stages of peptide–membrane interaction, particularly for cationic amphiphilic systems[51, 52]. These findings indicate that membrane charge and fluidity work together to regulate peptide insertion. Electrostatic attraction promotes the initial binding of the peptide, while membrane fluidity allows for structural adjustment and deeper insertion.

The mammalian and red blood cell (RBC) membrane models both showed complete exclusion of Alt-SNAPP from the bilayer interface, as reflected in the absence of overlap between the peptide and lipid density profiles as shown in Figures 4c and 4d. In the mammalian model, composed entirely of phosphatidylcholine (POPC), SNAPP remained fully separated from the membrane, with its density peaking well above the bilayer surface. This exclusion is potentially attributed to the zwitterionic nature of POPC and the lack of anionic headgroups, which are essential for electrostatic attraction of cationic peptides[44]. Furthermore, the zwitterionic nature of POPC and its higher lipid order further suppress nonspecific adsorption. Similarly, in the RBC membrane model characterized by its asymmetric leaflet composition, SNAPP was also fully excluded as shown in Figure 4d. The outer leaflet, rich in sphingomyelin (43 percent) and cholesterol (11 percent), presents a highly ordered and electrically neutral interface that resists peptide binding. Although the inner leaflet contains more phosphatidylethanolamine and phosphatidylserine, these anionic components are inaccessible from the solvent-facing side. Together, these findings are consistent with experimental reports showing that SNAPPs exhibit minimal cytotoxicity toward mammalian cells and extremely low haemolytic activity, likely due to their inability to interact with cholesterol and sphingomyelin enriched membrane regions[9, 47, 53].

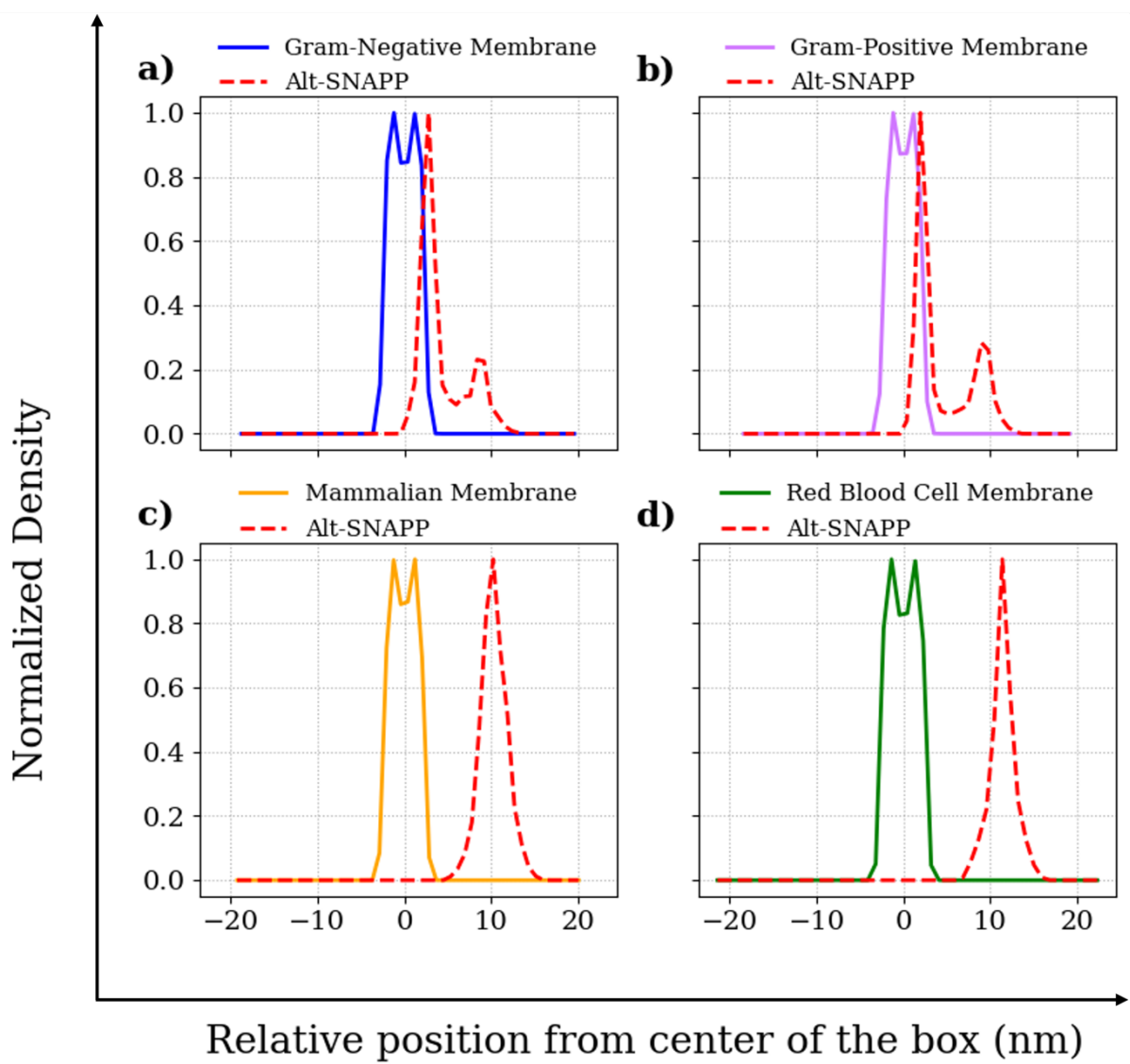

*Figure 4 - Partial density analysis of Alt-SNAPP and membrane along the bilayer normal for each simulated system: a) Gram-negative membrane, b) Gram-positive membrane, c) mammalian membrane, and d) red blood cell membrane.*

Together, the partial density profiles reinforce the membrane selectivity of SNAPP observed in COM distance analysis. SNAPP is preferentially enriched near anionic bacterial membranes and is excluded from zwitterionic and cholesterol dense mammalian and RBC membranes. The modestly greater insertion into the Gram-positive membrane compared to the Gram-negative system may reflect increased electrostatic stabilization due to higher anionic content, despite reduced fluidity. These results support the design rationale behind SNAPPs, emphasizing the role of lipid headgroup charge, membrane stiffness, and compositional asymmetry in mediating selective targeting of microbial membranes.

### 3.3. Hydrogen Bonding Interactions Between SNAPP Residues and Lipid Headgroups

To further investigate the molecular determinants of SNAPP's membrane selectivity, we analysed hydrogen bonding (H-bond) interactions between SNAPP residues and lipid headgroups across the four membrane models. This analysis provides atomic level insight into the specific interactions that stabilize SNAPP binding and contribute to its membrane affinity.

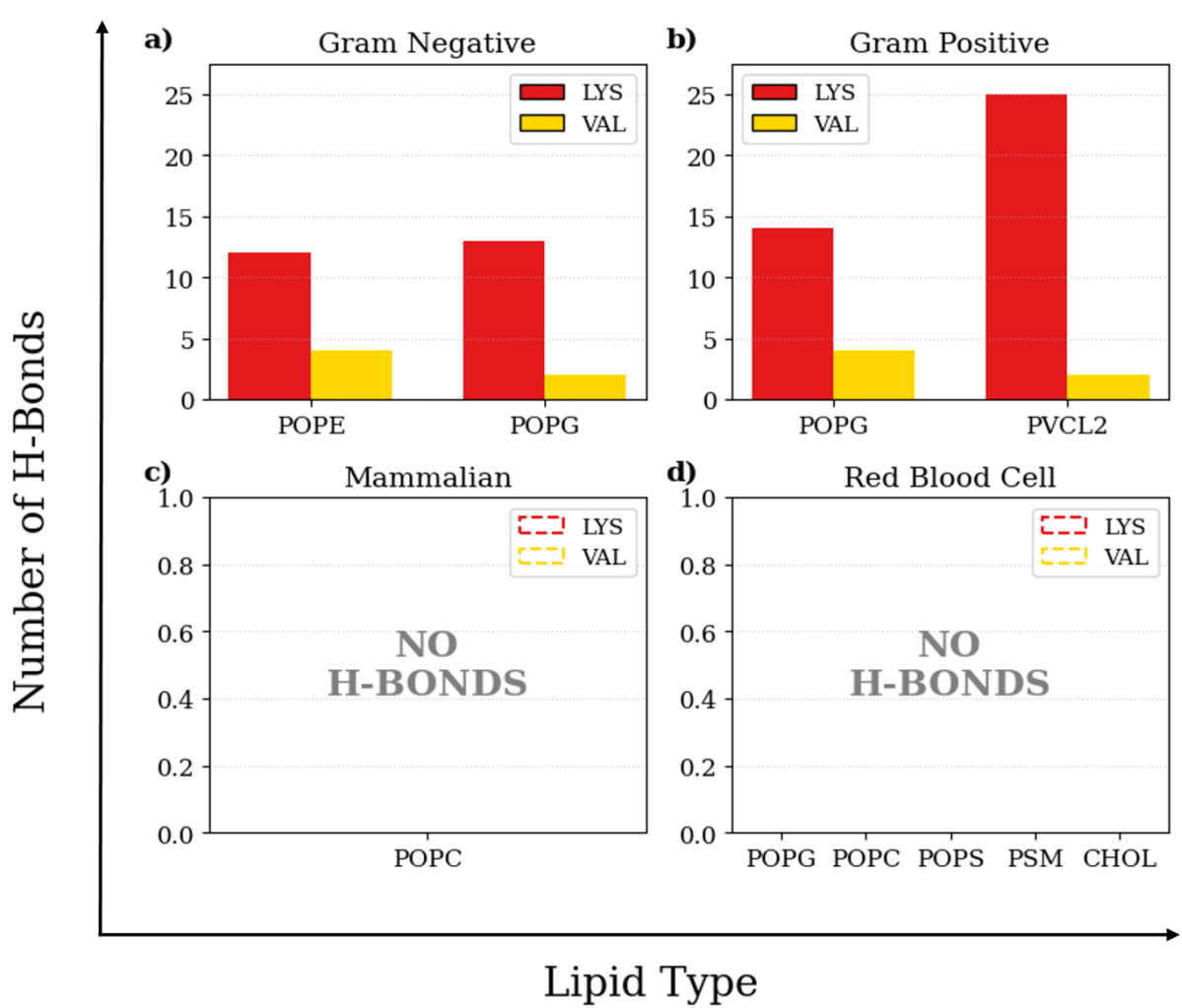

*Figure 5 - Hydrogen bond analysis between Alt-SNAPP residues (lysine and valine) and individual lipid types for each membrane system: a) Gram-negative, b) Gram-positive, c) mammalian, and d) red blood cell membranes.*

In the Gram-negative membrane as shown in Figure 5a, significant number of H-bonds were observed, particularly between lysine residues of SNAPP and both lipid types. Lysine formed approximately 12 to 13 H-bonds with POPE and POPG, respectively, while valine contributed a lower number of 3 to 4 H-bonds per lipid. These results are consistent with the known ability of primary amine groups in lysine side chains to donate hydrogen bonds to phosphate and carbonyl oxygen atoms in lipid headgroups[44, 51]. The modest contribution from valine reflects its nonpolar character and limited capacity for hydrogen bonding, suggesting that its role is likely confined to hydrophobic membrane interactions rather than polar anchoring.

In the Gram-positive membrane, as shown in Figure 5b, a similar hydrogen bonding pattern was observed as in the Gram-negative system, but with significantly greater magnitude. Lysine residues formed approximately 25 hydrogen bonds with PVCL2 and around 13 with POPG, indicating a dense and extensive interaction network. The increased number of hydrogen bonds in this system may result from greater electrostatic stabilization due to the higher overall density of anionic headgroups, as well as the increased conformational availability of lysine residues for polar contacts.

PVCL2, a cardiolipin species with a net charge of -2, enhances the membrane's negative surface potential and provides multiple phosphate and hydroxyl groups that act as effective hydrogen bond acceptors[54]. The deeper partial density profile of SNAPP observed in this system is consistent with these intensified polar interactions, supporting the view that electrostatic forces and hydrogen bonding cooperatively stabilize peptide insertion[55, 56].

Taken together, the hydrogen bond analysis highlights the central role of lysine residues in mediating membrane engagement through polar interactions, particularly in the context of anionic bilayers. The absence of hydrogen bonding in mammalian and RBC systems as shown in Figures 5c and 5d provides further molecular level validation of SNAPP's high specificity toward bacterial membranes and supports its potential as a selective and non-toxic antimicrobial agent.

### 3.4. Energetic Profile of Alt-SNAPP Binding to Different Membranes

### 3.4.1 Energetic Profile of Alt-SNAPP Binding to the Gram-negative Membrane

To gain quantitative insight into the molecular forces driving Alt-SNAPP interaction with Gram-negative bacterial membranes, we calculated the nonbonded interaction potential energies between Alt-SNAPP residues and the membrane components over the 500 ns simulation. The interaction energies were decomposed into Coulombic (electrostatic), van der Waals (Lennard-Jones), and total contributions, focusing on lysine and valine amino acids separately, followed by whole-complex SNAPP membrane interactions.

Figure 6a to 6d show the residue specific interaction energies for Gram-negative membrane system. Lysine exhibited consistently stronger interactions with both POPE (Figure 6a) and POPG (Figure 6c) compared to valine (Figures 6b and 6d). In both lipid environments, lysine reached total interaction energies more favourable than −3000 kJ/mol, dominated by Coulombic terms, indicating strong electrostatic attraction between lysine's protonated amine groups and the negatively charged phosphate and glycerol moieties of the lipid headgroups[37, 44]. POPG, having a higher net negative charge than POPE, facilitated even stronger lysine binding, reflected by lower interaction energy minima (reaching nearly −4000 kJ/mol) and a more stable binding profile. These results are consistent with prior studies showing that cationic antimicrobial polymers preferentially bind to PG-containing bilayers due to enhanced charge complementarity[57, 58].

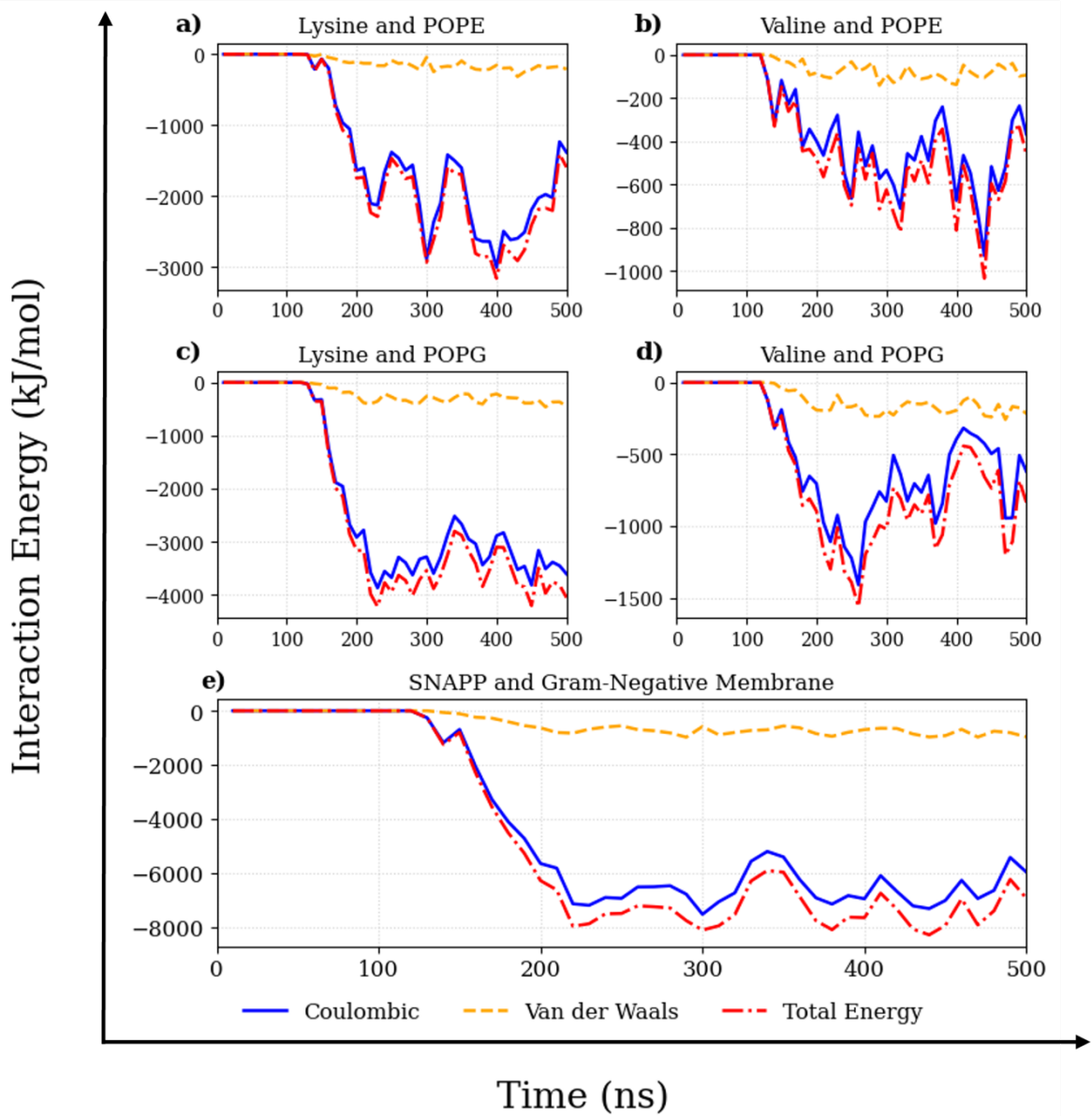

*Figure 6 -Nonbonded interaction potential energies between Alt-SNAPP residues and individual lipid types in the Gram-negative membrane system, showing Coulombic, van der Waals, and total energy components: a) lysine–POPE, b) valine–POPE, c) lysine–POPG, d) valine–POPG, and e) total interaction energy between Alt-SNAPP and the Gram-negative membrane.*

By contrast, valine formed weaker and more transient interactions with both lipid types, with total energies not exceeding −800 kJ/mol. The interaction energy curves for valine lipid pairs remained relatively flat, dominated by minor van der Waals contributions. These findings confirm valine's limited role in specific binding and suggest its primary function within Alt-SNAPP is to modulate amphiphilic balance and assist in membrane insertion via hydrophobic interactions[56, 59].

Figure 6e illustrates the total interaction energy between Alt-SNAPP and the entire Gram-negative bilayer. A sharp and sustained decrease in interaction energy begins around 150 ns, eventually reaching nearly −8000 kJ/mol by the end of the simulation. This energetic stabilization is primarily driven by Coulombic contributions, confirming that electrostatic interactions govern Alt-SNAPP's affinity toward the Gram-negative membrane. The time window of energetic stabilization closely mirrors the reduction in center-of-mass distance and the partial density overlap observed in earlier analyses, indicating synchronized spatial and energetic convergence. This energetics are characteristic of multivalent polymer membrane interactions observed in experimental studies of cationic antimicrobial polymers[9, 56].

Together, these results highlight the importance of residue specific electrostatics in dictating the binding strength of Alt-SNAPP to anionic bacterial membranes. The energetic dominance of lysine, particularly in interaction with PG lipids, underscores its central role in membrane targeting. This energetic basis reinforces the broader mechanistic model in which electrostatic attraction, and not just hydrophobicity or passive partitioning, is the dominant driver of bacterial membrane selectivity in Alt-SNAPP systems.

### 3.4.2 Energetic Profile of SNAPP Binding to Cardiolipin-Enriched Gram-Positive Membranes

Building upon the interaction energy analysis of Alt-SNAPP with the Gram-negative bilayer, we next examined the energetic landscape of SNAPP association with a Gram-positive membrane.

As shown in Figures 7a and 7c, lysine residues displayed strong and progressively stabilizing interactions with both POPG and cardiolipin over the course of the 500 ns simulation. The lysine and POPG interaction energy reached approximately −4000 kJ/mol by the end of the trajectory, while interaction between lysine and cardiolipin was particularly strong, with nonbonded interaction potential energies descending below −6000 kJ/mol. These values were predominantly driven by Coulombic forces, reflecting the high charge density of both the doubly deprotonated cardiolipin headgroup and the fully protonated lysine side chains. Cardiolipin's enhanced interaction with lysine can be attributed to its bulky, branched polar headgroup, which presents multiple phosphate and hydroxyl groups acting as hydrogen bond acceptors. This structural arrangement enables multivalent binding and strengthens electrostatic stabilization, making cardiolipin a key contributor to the observed interaction landscape[37, 55]. This observation supports earlier findings from the hydrogen bond analysis, which indicated a high number of lysine mediated hydrogen bonds with cardiolipin.

In contrast, valine interactions with both lipids were minimal (Figures 7b and 7d), with total interaction energies stabilizing around −800 to −1000 kJ/mol. These energies were dominated by van der Waals contributions and reflect nonpolar side chain contacts with lipid acyl chains. The functional role of valine in this context remains amphiphilic, contributing to the hydrophobicity of the polymer without engaging in specific interactions[56, 59, 60].

The total interaction energy between SNAPP and the Gram-positive membrane (Figure 7e) further confirmed the strong affinity of SNAPP for cardiolipin rich bilayers. After an initial lag phase, the interaction energy declined sharply after ~180 ns, stabilizing near −11,000 kJ/mol, significantly more favourable than the −8000 kJ/mol observed for the Gram-negative membrane. This deeper energetic well suggests that once SNAPP overcomes the initial kinetic barrier to association, its interaction with cardiolipin enriched membranes is highly stable and energetically advantageous. This correlates with the deeper SNAPP insertion observed in the partial density profiles and the higher frequency of hydrogen bonding in this system.

Taken together, these results highlight the central role of lysine and lipid electrostatic interactions, particularly with cardiolipin, in mediating strong and selective SNAPP binding to Gram-positive membranes. While Gram-negative membranes promote faster initial association due to higher fluidity, the cardiolipin rich Gram-positive bilayer provides a more electrostatically robust and thermodynamically stable platform for SNAPP anchoring. These findings are consistent with previous reports that antimicrobial peptides and polymers preferentially accumulate in cardiolipin rich microdomains of bacterial membranes[44, 57], and they reinforce the potential of cardiolipin as a key molecular target for selective antibacterial therapies.

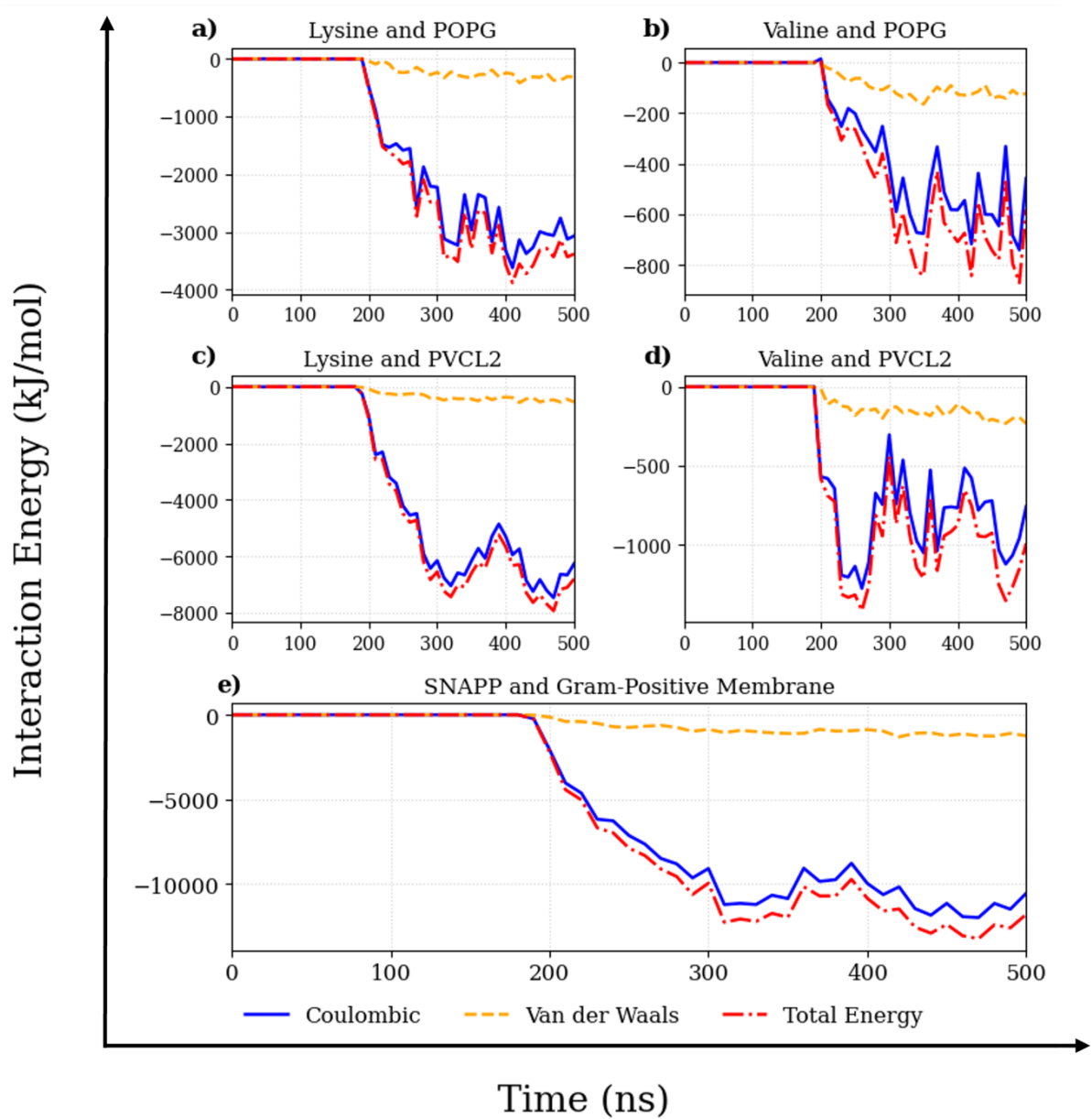

*Figure 7 - Nonbonded interaction potential energies between Alt-SNAPP residues and individual lipid types in the Gram-positive membrane system, showing Coulombic, van der Waals, and total energy components: a) lysine–POPG, b) valine–POPG, c) lysine–PVCL2, d) valine–PVCL2, and e) total interaction energy between Alt-SNAPP and the membrane.*

## 4.4.3. Absence of Interaction Energy Confirms SNAPP Exclusion from Mammalian and Red Blood Cell Membranes

For the mammalian system (Figure S1), energy decomposition at the residue level for lysine with POPC (Figure S1a) and valine with POPC (Figure S1b), as well as for the entire Alt-SNAPP membrane complex (Figures 8 and S1c), revealed that all components of the interaction energy remained at absolute zero throughout the 500 ns trajectory. There were no observable fluctuations or transient values in either electrostatic or dispersion forces. This strongly supports the conclusion that SNAPP remains completely excluded from the POPC bilayer and does not engage in surface binding or insertion.

In the RBC membrane system (Figure S2), energy was decomposed for interactions between SNAPP and ten distinct lipid residue pairs (Figures S2a–S2j), as well as the total Alt-SNAPP membrane interaction energy (Figures 9 and S2k). As with the mammalian case, all energies Coulombic, van der Waals, and total remained zero across the entire trajectory, regardless of the lipid species or residue type. Even interactions with phosphatidylserine (POPS), which carries a net negative charge and resides in the inner leaflet, produced no detectable energetic signature. This likely reflects the spatial inaccessibility of POPS from SNAPP's solvent-facing position and the protective role of the outer leaflet, which is dominated by zwitterionic sphingomyelin (PSM), POPC, and membrane-stiffening cholesterol[47]. These components produce a tightly packed, electrostatically neutral interface that resists peptide adsorption, a well-documented characteristic of erythrocyte membranes[45].

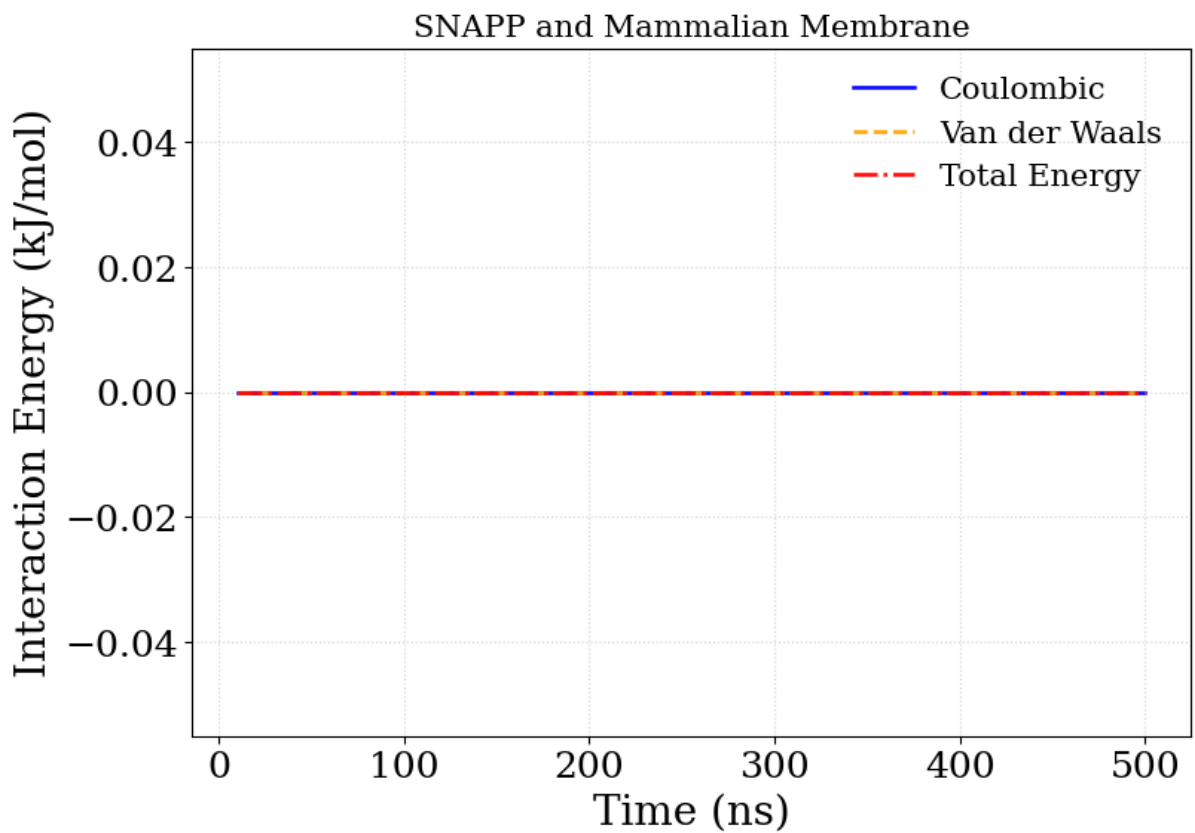

*Figure 8 -Nonbonded total interaction energy between Alt-SNAPP and the mammalian membrane.*

The complete absence of interaction energy in both host-derived systems contrasts starkly with the substantial energy drops observed in the Gram-negative and Gram-positive simulations, where Coulombic energy alone reached magnitudes of −8000 to −11,000 kJ/mol due to extensive interactions with POPG and cardiolipin. In the mammalian and RBC systems, the lack of even transient energetic contact reinforces prior observations from the center-of-mass and partial density analyses, where SNAPP was shown to remain distant from the bilayer surface and completely solvated in bulk water.

These results provide conclusive energetic validation of SNAPP's membrane-selective mechanism, in which bacterial membranes serve as attractive, energetically favourable targets due to their high anionic lipid content, while host membranes present electrostatically neutral, physically rigid barriers to interaction. The absence of measurable nonbonded interaction energy across all components, supported by the flat line energy profiles, confirms that SNAPP does not engage with mammalian or RBC membranes under physiologically relevant conditions. This energetic neutrality underpins SNAPP's low haemolytic potential and its capacity to discriminate between bacterial and host cells, a key design criterion for clinical viability[9].

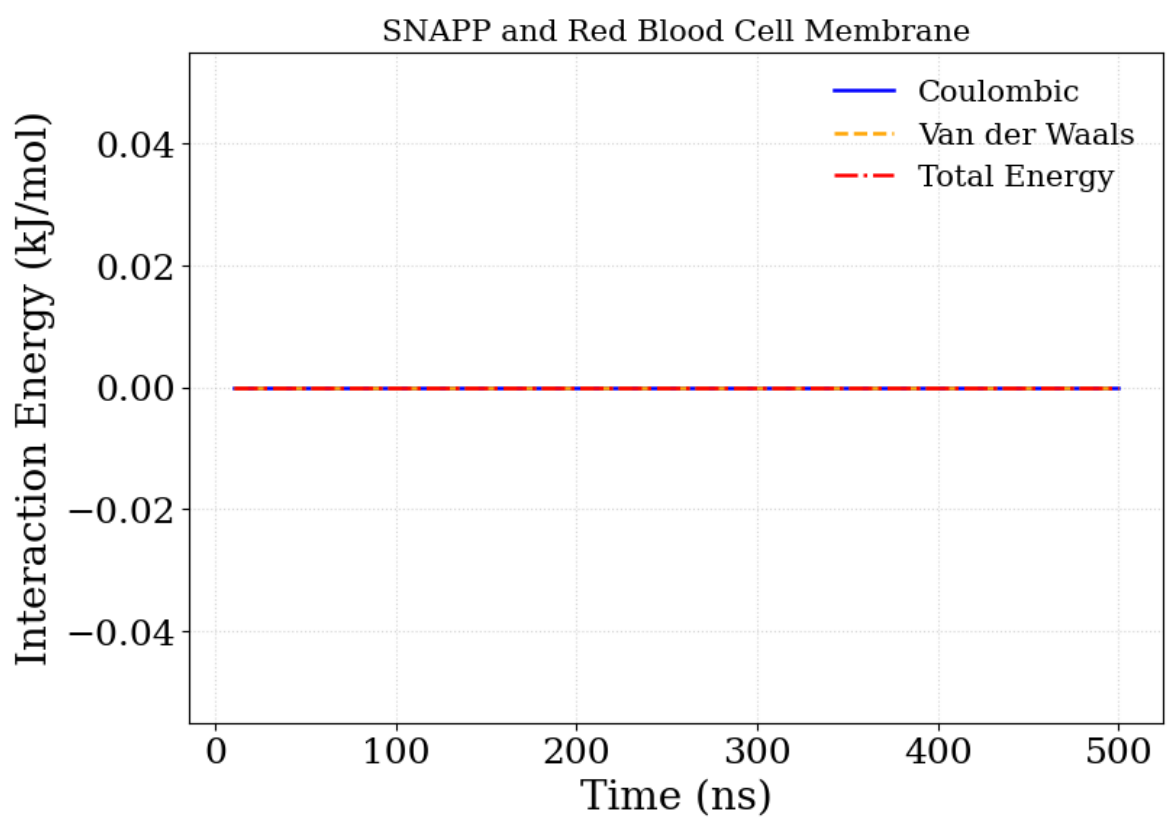

*Figure 9 - Nonbonded total interaction energy between Alt-SNAPP and the red blood cell membrane.*

## 4. Conclusion

This study provides an atomistic perspective on the selective membrane interaction behaviour of Alt-SNAPP, a novel star shaped anti-microbial peptide polymer, across four representative lipid bilayer models, representing Gram-negative, Gram-positive, mammalian, and red blood cell membranes. Through a comprehensive suite of molecular dynamics simulations, including center of mass distance tracking, partial density profiling, hydrogen bond analysis, and nonbonded interaction potential energy decomposition, we identified the molecular determinants that govern SNAPP's preferential association with bacterial membranes and its exclusion from host derived bilayers.

Our findings demonstrate that Alt-SNAPP rapidly and stably associates with Gram-negative membranes due to the strong electrostatic attraction to phosphatidylglycerol (PG) and enhanced accessibility from the fluid and loosely packed nature of phosphatidylethanolamine (PE). In Gram-positive membranes containing both phosphatidylglycerol (PG) and cardiolipin (CL), SNAPP displayed deeper insertion and more favourable interaction energies, primarily through multivalent electrostatic and hydrogen bonding interactions. Cardiolipin (CL), with its high negative charge and branched polar headgroup, played a particularly important role in stabilizing SNAPP engagement.

In contrast, mammalian and red blood cell membranes composed predominantly of zwitterionic lipids such as phosphatidylcholine (PC) and sphingomyelin (SM), along with elevated cholesterol content, exhibited complete exclusion of SNAPP. These bilayers presented highly ordered, electrostatically neutral surfaces that resisted peptide binding, as confirmed by the absence of hydrogen bonds, partial density overlap, and nonbonded interaction energy across the simulation timeframe.

Together, these results provide detailed molecular evidence that SNAPP's membrane selectivity is governed by the electrostatic complementarity between cationic residues and anionic lipids, particularly phosphatidylglycerol (PG) and cardiolipin (CL). This selectivity mechanism is consistent with the experimental data showing minimal cytotoxicity and haemolytic activity[9]. These findings support the broader design principle that optimizing cationic and amphiphilic balance enables targeting of bacterial membranes while avoiding host membrane disruption. This insight provides a foundation for the future development of selective antimicrobial polymers with therapeutic potential and reduced off target effects using active[61-63] and explainable[64-66] machine learning and advanced optimization techniques[67-68].

**Acknowledgment**

A.J is supported by research training programme (RTP) scholarship provided by the Australian government. This research was supported by The University of Melbourne's Research Computing Services and the Petascale Campus Initiative.

This research was partially funded by the Australian Government through the Australian Research Council and E.H and G.Q acknowledge the financial support through the ARC Discovery Project grant DP250101065.

**Conflict of interest**

Authors declare no conflict of interest.